# Modelling quantum photonics on a quantum computer

Anton N. Vetlugin[1*], Cesare Soci[1] and Nikolay I. Zheludev[1,2]

[1]*Centre for Disruptive Photonic Technologies, SPMS, TPI, Nanyang Technological University, Singapore 637371*
[2]*Optoelectronics Research Centre and Centre for Photonic Metamaterials, University of Southampton, Southampton SO17 1BJ, United Kingdom*
*Corresponding author. Email: a.vetlugin@ntu.edu.sg (A.N.V)

**Modelling of photonic devices traditionally involves solving the equations of light-matter interaction and light propagation, and it is restrained by their applicability. Here we demonstrate an alternative modelling methodology by creating a "quantum copy" of the optical device in the quantum computer. As an illustration, we simulate quantum interference of light on a thin absorbing film. Such interference can lead to either perfect absorption or total transmission of light through the film[1-3], the phenomena attracting attention for data processing applications in classical[4,5] and quantum[6] information networks. We map behaviour of the photon in the quantum interference experiment to the evolution of a quantum state of transmon[7,8], a superconducting charge qubit of the IBM quantum computer[9]. Details of the real optical experiment are flawlessly reproduced on the quantum computer. We argue that superiority of the "quantum copy" methodology shall be apparent in modelling complex multi-photon optical phenomena and devices.**

Quantum computing primarily aims to tackle computational problems such as integer factorization and unstructured search[10]. Beyond this, quantum computers provide a remarkable opportunity to simulate behaviour of quantum systems[11] with emphasis on quantum chemistry, quantum materials, particle physics and cosmology[12]. Here, we are interested in application of quantum computing to simulation of phenomena of quantum photonics and devices based on the first principles, without relying on the fundamental equations. The core of the methodology is the creation of a "quantum copy" of a real optical experiment by developing a simulator-system correspondence, Fig. 1. To provide an illustrative example, we program the IBM quantum computer[9] to implement a quantum copy of the effect of single photon interference on a thin absorber. This phenomenon is of great interest for fundamental research[2,13,14] as well as for practical applications in light harvesting, detection[15], sensing[16] and photonics data processing[4,5,17].

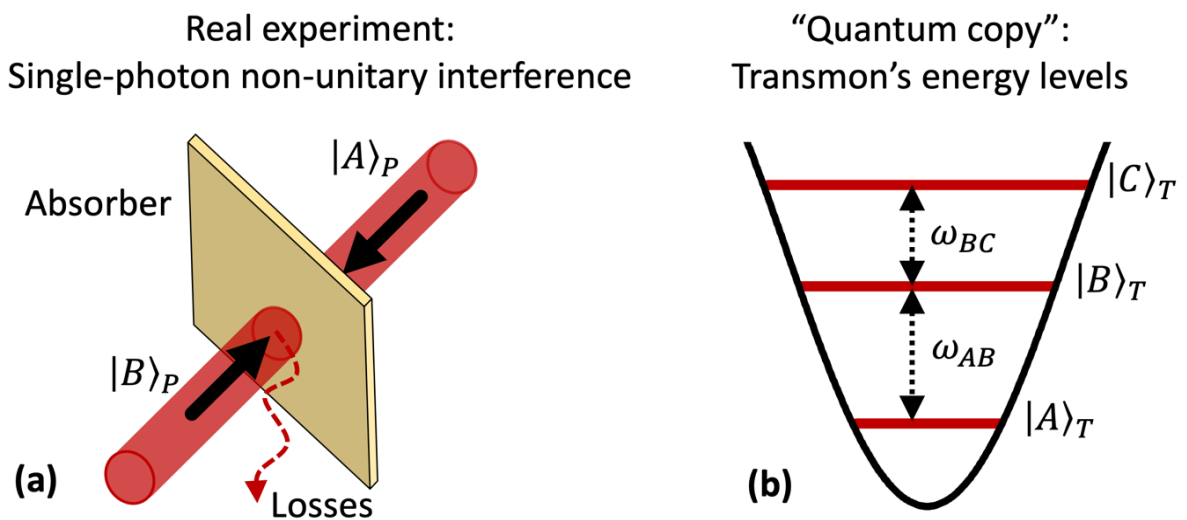

Figure 1. **Building a "quantum copy" of the photon interference experiment on a transmon**. A correspondence is established between the optical interference experiment with a thin absorber (a) and a transmon of a quantum computer (b): The presence of a photon in the input ports $|A\rangle_P$ and $|B\rangle_P$ corresponds to states $|A\rangle_T$ and $|B\rangle_T$ of the trasmon. Losses are represented by the transition to the excited state $|C\rangle_T$.

In classical optics, interference of coherent light waves on a thin absorber can eliminate the Joule losses of light energy or can lead to the total absorption of light depending on the phase relationship of the interfering waves[1]. Interference on a thin absorbing film was also extensively studied in quantum optics[3,6,18-25]. For instance, it was demonstrated that, despite the probabilistic nature of single photon absorption in a travelling wave, the regimes of deterministic absorption and deterministic transmission of a photon can be achieved in a standing wave[3]. Below, we show how to model this experiment on a quantum computer.

A sketch of the setup, used in the experiment of Ref. 3, is shown in Fig. 2(a). The photon enters interferometer through the 1:1 beamsplitter BS (step 1) where it can be either transmitted to one or the other arms of interferometer. The phase delay $\varphi$ between interferometer's arms is used to control the inner phase of a single photon wavefunction (step 2), so in the bra-ket notation propagation through interferometer can be written as:

$$|A\rangle_P \to \frac{1}{\sqrt{2}}\bigl(|A\rangle_P + e^{i\varphi}|B\rangle_P\bigr). \tag{1}$$

where $|A\rangle_P$ and $|B\rangle_P$ identify the path of the photon, Fig. 2(a). This wavefunction interferes on a thin absorber (a plasmonic metamaterial film of a subwavelength thickness in the experiment) placed in the middle of the interferometer (step 3). The absorber is designed to have equal reflection and transmission coefficients of 25% and traveling wave absorption of 50%. The absorber becomes totally transparent if the photon is prepared in the anti-symmetric state, $(|A\rangle_P - |B\rangle_P)/\sqrt{2}$, $(\varphi = \pi)$. In contrast, the symmetric state of the photon, $(|A\rangle_P + |B\rangle_P)/\sqrt{2}$, $(\varphi = 0)$ is deterministically dissipated by the absorber. Indeed, the antisymmetric state corresponds to a standing wave in the interferometer with its node (zero electric field) coincident with the absorber's position[25]: the photon escapes dissipation due to the negligible electric field at the absorber. In contrast, the symmetric state corresponds to a standing wave with its anti-node (maximum electric field) at the absorber's position where light dissipation reaches 100%. For arbitrary $\varphi$, redistribution of the photon's wavefunction between the symmetric and anti-symmetric states defines the probability of the photon absorption and transmission as $\cos^2\frac{\varphi}{2}$ and $\sin^2\frac{\varphi}{2}$, correspondingly[18,25] (step 4). In the experiment, the probability of photon transmission is measured by sending a stream of photons through the interferometer and comparing the number of detected photons with and without the absorbing film. By changing the phase of the interferometer $\varphi$, the transition between the regimes of "perfect absorption" and "perfect transmission" is observed, Fig. 3(a).

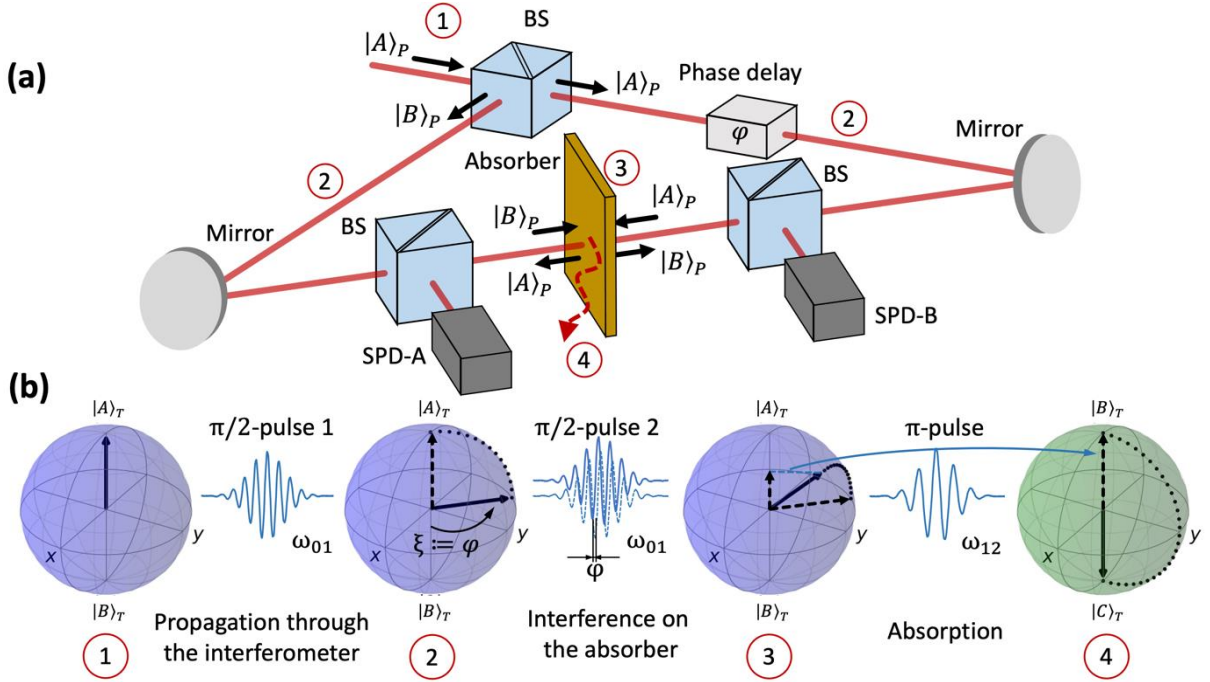

Figure 2. **Mapping evolution of a single photon in a lossy interferometer to the transmon's dynamics.** (a) Photon enters the interferometer consisting of the input beamsplitter (BS), phase delay and mirrors. A thin absorber (plasmonic metamaterial) is placed in the middle of the interferometer. The outgoing light is detected by single photon detectors SPD-A and SPD-B. Steps of the photon's evolution are indicated by the numbers in circles. (b) The dynamics of the transmon is shown on the Bloch spheres representing its quantum state for the different stages of photon's evolution: ①→② A $\pi/2$-pulse at $\omega_{AB}$ changes the stage of transmon to mimic propagation of the photon through the interferometer. ②→③ The second $\pi/2$-pulse, also at $\omega_{AB}$, but phase shifted on $\varphi$ changes the state of the transmon to mimic interference on the absorber. ③→④ The action of a $\pi$-pulse at $\omega_{BC}$ mimics photon's absorption by moving the transmon to the higher excited state $|C\rangle_T$.

We have built a quantum copy of this experiment on the IBM quantum computer with the ibmq_armonk[9] processor. The key functional element of the processor is a transmon[7] ("transmission-line shunted plasma oscillation qubit"), a type of superconducting charge qubit built around two Josephson junctions shunted with an additional capacitor. The transmon is an anharmonic quantum oscillator with a non-equidistance energy spectrum that allows the transition between two given states to be isolated and selectively addressed. We denote the ground, first excited and second excited states ("basis" states) of the transmon as $|A\rangle_T$, $|B\rangle_T$ and $|C\rangle_T$, respectively, where the frequency of transitions $|A\rangle_T \to |B\rangle_T$ and $|B\rangle_T \to |C\rangle_T$ are $\omega_{AB} \approx 4.97$ GHz and $\omega_{BC} \approx 4.62$ GHz, respectively.

Control of the transmons is performed by microwave signals from coupled "drive" resonator[8]. For instance, in the oscillatory driving field at the resonant frequency $\omega_{AB}$, the transmon will undergo the cyclic Rabi behaviour, oscillating between the ground and first excited states. A microwave pulse transferring the transmon from the ground to the first excited state is known as a π-pulse, while a π/2-pulse leaves the transmon halfway through, in a superposition of the ground and first excited states. A sequence of such control pulses is used in the modelling to mimic different stages of evolution of light quantum in the interferometer. Here it is important that the transmon's longitudinal and transverse relaxation times are of the order of few tens of $\mu$s[9] which is much longer than the experimental cycle that involves consecutive application of three control microwave pulses of 0.6 $\mu$s each (see details below): natural relaxation of the transmon's excited states is insignificant during the experimental cycle. The energy state of the transmon is measured using a coupled "readout" microwave resonator: the magnitude and phase of the probe signal reflected from the "readout" resonator identify the transmon's state using a state discriminator[26,27] calibrated on the basis energy states prior to the experiment.

Assuming other states are unpopulated, the state of the transmon in the basis of $|A\rangle_T$ and $|B\rangle_T$ is depicted as a vector on the Bloch sphere[10] (purple spheres in Fig. 2(b)) where the north and south poles of the sphere represent the states $|A\rangle_T$ and $|B\rangle_T$, respectively. Any transformation of the transmon state can be shown as a rotation(s) on the Bloch sphere. On the transmon, propagation through the interferometer and redistribution of the photon's wavefunction at the absorber is mimicked by driving the transmon with resonant microwave pulses at frequency $\omega_{AB}$. Initiating the transmon in its ground state $|A\rangle_T$ corresponds to sending the photon through the input port of the interferometer (step 1 in Fig. 2). The transmon's state transformation, mimicking photon's propagation through the interferometer from step 1 to step 2, equation (1), is induced by a $\pi/2$-pulse at $\omega_{AB}$. Redistribution of the photon's wavefunction between symmetric and anti-symmetric states in the middle of the interferometer (step 3) is replicated on the transmon by the second $\pi/2$-pulse at $\omega_{AB}$ with a phase shift $\varphi$ compared to the first pulse. Here, the result of the modelling depends only on the phase difference between two consecutive $\pi/2$-pulses and not on their absolute phases. The second $\pi/2$-pulse rotates the anti-symmetric component of the wavefunction to the ground state and the symmetric component to the first excited state. To mimic absorption of the symmetric component of the photon's wavefunction (step 4), a $\pi$-pulse at $\omega_{BC}$ is applied which brings the population of the first excited state of the transmon to the higher excided state $|C\rangle_T$. The $\pi$-pulse at $\omega_{BC}$ is off-resonant for the $|A\rangle_T \to |B\rangle_T$ transition and the population of the ground state, corresponding to photon's transmission through the absorber, is not affected.

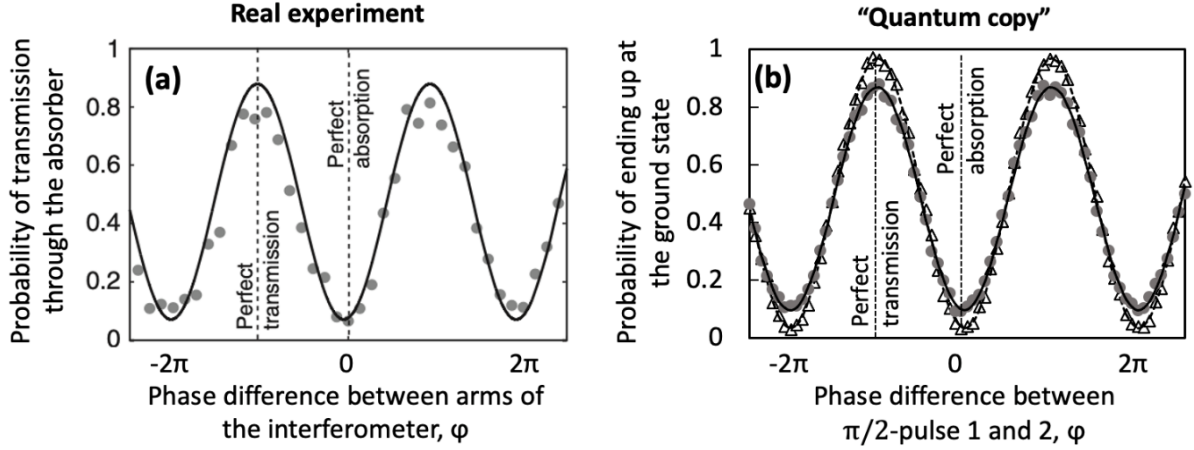

Figure 3. **A comparison between the outcomes of a real experiment of quantum interference on a thin absorber and a modelling experiment on a quantum computer.** (a) The probability of a single photon transmission through the absorber is a function of $\varphi$, the phase difference between the arms of interferometer. Note the regimes of perfect absorption ($\varphi = 0$) and perfect transmission ($\varphi = \pm\pi$). (b) The probability of photon's transmission is represented by the probability of finding transmon in the ground state: modelling for the optimized absorber (empty triangles and dashed fitting line) and absorber with parameters close to the experiment (grey circles and solid fitting line). This probability depends on the phase delay $\varphi$ between the first and second $\pi/2$-pulses acting on the transmon. This phase delay represents the optical pass phase difference between the arms of interferometer.

In the optical experiment, a stream of single photons is sent through the interferometer with a fixed phase delay $\varphi$ to find the probability of photon's transmission by measuring a large number of single photon events, Fig. 3(a). To mimic this on the quantum computer, we run a number of trials (~1000) with a fixed phase difference $\varphi$ between the first and second $\pi/2$-pulses. The probability of the transmon to end up in the ground "transmitted" is presented in Fig. 3(b).

Fig. 3(b) shows the results of quantum interference modelling for absorber optimized to achieve a perfect absorption and for absorber with parameters close to the real experiment. The modelling accurately reproduces real experiment exhibiting the regimes of "total absorption" and "total transmission" depending on the phase $\varphi$. Nearly perfect visibility of the simulated dependence of absorption probability on $\varphi$ is achieved with optimised absorber (50% traveling wave absorption). In the experiment of Ref. 3 the metamaterial's absorption was 45%, i.e. off the optimized value. That explains lower visibility observed in the real experiment that shows some residual absorption for $\varphi = \pm\pi$ and non-zero transmission for $\varphi = 0$ due to unperfect interference cancellation on the absorber, Fig.3(a). To model the non-optimised absorber the amplitude of the second pulse controlling transmon was decreased by about 30% from the amplitude of the $\pi/2$-pulses. Such adjustment of the control pulse leads to the absorption increase of the anti-symmetric component and corresponding absorption decrease of the symmetric component of the transmon's wavefunction mimicking interference disbalance in the real experiment.

Modelling of optical phenomena on quantum computer can be developed further in different ways. By incorporating higher excited states, one can consider various models of losses and non-unitary transformation, incorporate the mechanisms of radiative and non-radiative decay etc. At the same time, employment of higher excited states would allow to study propagation of a single photon through a more complex networks in the form of the multi-port (non-unitary) interferometers[28,29] and quantum memories[30] where the correspondence between the presence of quantum light in different ports of the network and energy states of the transmon can be established. By exploiting multiple transmons, dynamics of different quantum states of light[25], such as NOON[19] and entangled[23] states, can be studied, and this regime would be of high interest since optical experiments on the generation and detection of complex quantum states of light is a challenging task. Similarly, other optical phenomena can be modelled by building a simulator-system correspondence. For instance, using mathematical identity of the Bloch and Poincare sphere, quantum copy of experiments involving multidimensional structured quantum light[31] can be developed.

In conclusion, the methodology of building a quantum copy of the real optical experiment provides a universal platform for studying and analysis of quantum optics phenomena and systems. We argue that modelling on a quantum computer with a discrete spectrum of transmon is the next, quantum step of developing analogue computing in photonics since the wide-spread use of classical analogue computers to solve problems of nonlinear optics in the 1970s[32].

**Acknowledgement**

The authors are grateful to IBM for free access to services provided by the IBM Quantum Researchers Program. The views expressed are those of the authors, and do not reflect the official policy or position of IBM or the IBM Quantum team. This work was supported by the Singapore Ministry of Education (MOE2016-T3-1-006 (S)), the Singapore NRF-Quantum Engineering Program (NRF-QEP1) and the UK's Engineering and Physical Sciences Research Council (grant EP/M009122/1).

**Author contributions**

A.N.V conceived the idea and performed the modelling experiment; A.N.V. and N.I.Z. wrote the manuscript; all co-authors discussed the results. N.I.Z. & C.S. co-supervised the project.